\documentclass[namedreferences]{solarphysics}
\usepackage[optionalrh]{spr-sola-addons} % For Solar Physics 
\usepackage{graphicx}        % For eps figures, newer & more powerfull
\usepackage{enumerate}        
\usepackage{color}           % For color text: \color command
\usepackage{url}             % For breaking URLs easily trough lines
\newcommand{\solphys}{{\it Solar Phys.}}

\begin{document}

\begin{article}

\begin{opening}

\title{
Helioseismic Holography of an Artificial Submerged Sound Speed Perturbation and
Implications for the Detection of Pre-Emergence Signatures of Active Regions
}

\author{D.~C.~\surname{Braun}$^{1}$
       }
\runningauthor{D. Braun}
\runningtitle{Helioseismic Holography of an Artificial Perturbation}

\institute{$^{1}$ NWRA, CoRA Office, 3380 Mitchell Ln, Boulder CO 80301 USA
                     email: \url{dbraun@nwra.com} \\ 
             }

\begin{abstract}
We use a publicly available numerical wave-propagation simulation of 
\citeauthor{Hartlep2011} (\solphys{} \textbf{268}, 321, \citeyear{Hartlep2011})
to test the ability of helioseismic holography to detect signatures of a
compact, fully submerged, 5\,\% sound-speed perturbation 
placed at a depth of 50 Mm within
a solar model. We find that helioseismic holography as employed in a
nominal ``lateral-vantage'' or ``deep-focus'' geometry employing quadrants of an 
annular pupil is capable of detecting and characterizing the perturbation. 
A number of tests of the methodology, including 
the use of a plane-parallel approximation, 
the definition of travel-time shifts,
the use of different phase-speed filters, and changes to 
the pupils, are also performed. 
It is found that travel-time shifts made using 
Gabor-wavelet fitting are essentially
identical to those derived from the phase of the Fourier transform
of the cross-covariance functions.  The errors in travel-time
shifts caused by the plane-parallel 
approximation can be minimized to less than a second for the depths and
fields of view
considered here.  Based on the measured strength of the
mean travel-time signal of the perturbation, no substantial improvement in
sensitivity is produced by varying the analysis procedure from the 
nominal methodology in conformance with expectations. 
The measured travel-time shifts are essentially unchanged by varying
the profile of the phase-speed filter or omitting the filter entirely. 
The method remains maximally sensitive when
applied with pupils that are wide quadrants, as opposed to narrower
quadrants or with pupils composed of smaller arcs.
We discuss the significance of these results for 
the recent controversy regarding suspected pre-emergence signatures 
of active regions.
\end{abstract}
\end{opening}

%-------------------------------------------------

\section{Introduction}
     \label{S-Introduction} 

For almost two decades,
helioseismic methods have been employed to search for evidence of magnetic flux 
rising through the convection zone
\cite{Braun1995b,Chang1999,Jensen2001,Zharkov2008,Kosovichev2009,Hartlep2011,Ilonidis2011,Leka2012,Birch2012}.
Submerged magnetic fields may produce travel-time anomalies due to changes
in the wave speed caused by the magnetic field or by the presence of flows
and perturbations to the thermal structure associated with the magnetic 
field \cite{Birch2010}.
If positively identified, such signatures could play an important role in
space-weather forecasting, and lead to physical understanding of the emergence
process, which is a key component of the solar activity cycle.
Recent detection of $p$-mode travel-time anomalies prior to the emergence
of several large active regions, obtained with time--distance methods, have
been reported (\citeauthor{Ilonidis2011} 
\citeyear{Ilonidis2011}, \citeyear{Ilonidis2012a}) 
No significant travel-time anomalies were subsequently 
measured from an independent analysis using helioseismic holography 
\cite{Braun2012b}. \inlinecite{Ilonidis2012b} suggest that this discrepancy
may be due to differences in sensitivity between the methods employed.

Numerical simulations have provided artificial data through which 
helioseismic analysis and modeling can be tested 
\cite{Jensen2003c,Benson2006,Hanasoge2006,Parchevsky2007,Zhao2007,Braun2007,Cameron2008,Parchevsky2009,Crouch2010,Cameron2011,Birch2011,Hartlep2011,Braun2012}.
Many of these simulations include near-surface flows, sound-speed
perturbations, or magnetic structures typical of active regions or supergranulation.
Simulations that propagate waves through completely submerged perturbations
are rarer \cite{Hartlep2011}, but are critical for testing and developing 
helioseismic methods sensitive to looking for active regions prior to their
emergence on the surface. In this work, we use one of the simulations 
of \inlinecite{Hartlep2011} to test the sensitivity of helioseismic holography
comparatively to subsurface sound-speed perturbations under a variety 
of applications.

\section{Simulation} \label{S-Simulation}      

\begin{figure}
\centerline{\includegraphics[width=1.0\textwidth]{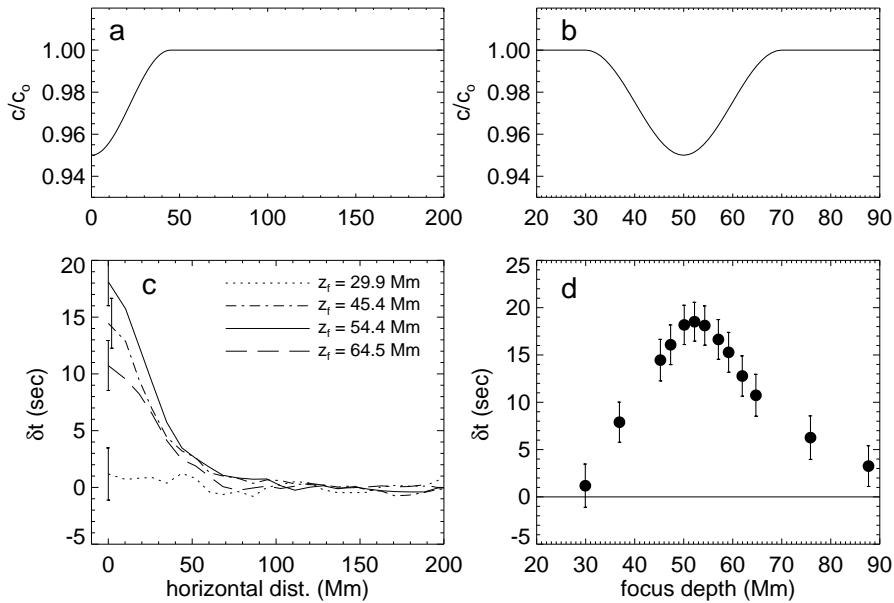}}
\caption{
The sound-speed ratio [$c/c_o$] in the simulation, where $c$ is the perturbed sound
speed and
$c_o$ is the background sound speed of Model S (top panels), and travel-time
measurements (bottom panels) made with helioseismic holography using a 
``nominal'' methodology (see text).   
(a) The variation with horizontal distance from the center of the circularly
symmetric sound-speed perturbation at a depth of 50 Mm below the surface 
of the simulation.
(b) The variation with depth of the sound-speed ratio
at the center of the perturbation. (c) The variation with horizontal
distance of the azimuthally averaged mean travel-time shift measured 
using lateral-vantage helioseismic
holography  applied to the simulation at focus depths of 29.9 Mm (dotted line),
45.4 Mm (dash--dot line), 54.4 Mm (solid lines), and 64.5 Mm (long dashed line). 
The travel-time shifts are averaged over 0.7$^\circ$-wide annuli
centered on the location of the perturbation. (d) The travel-time shift
at the center of the perturbation as a function of
focus depth. The error bars in panels (c) and (d) indicate the standard 
deviation of the realization
noise determined from a region away from the perturbation (see text).
}
\label{fig.nom1}
\end{figure}

\inlinecite{Hartlep2011} constructed a
number of simulations containing $p$ modes propagating through a
spherical domain containing localized perturbations of the sound speed
about the standard solar Model S \cite{JCD1996}. No flows or magnetic fields
are included. The solar model is convectively stabilized by a neglect of 
the entropy gradient of the background model, which lowers the acoustic
cut-off frequency. The mode amplitudes above 3.5 mHz are thus reduced in
amplitude. In addition, the simulation is only populated with $p$ modes 
with angular degree $\ell$ between 0 and 170. The simulations span 
about 17 hours of solar time. A number of simulations using the same code are publicly
available and include a variety of sound-speed perturbations at different
depths.  In this work, we employ the simulation with a peak 5\,\% sound-speed reduction
at a depth of 50 Mm and with a horizontal size of 45 Mm (see Figure~\ref{fig.nom1}).
The simulated velocity field is provided in arbitrary units and represented in 
heliographic coordinates,
with 512 pixels in longitude and 256 pixels in latitude, a cadence of one minute, 
and it is stored in a FITS file (see \url{sun.stanford.edu/~thartlep/Site/Artificial_Data/Entries/2012/3/21_Subsurface_sound_speed_perturbations.html}).

Our primary emphasis is on testing the ability of helioseismic holography
to detect $p$-mode travel time signatures of the prescribed perturbation
within the simulation, and to measure the {\it relative} sensitivity of the 
results (in both signal strength and background noise) to changes of methodology.
In contrast, direct comparisons of measured and expected travel-times 
requires the computation and application of sensitivity functions which is not
attempted here.  A prediction of
the travel-time shift expected from a given sound-speed perturbation 
is a non-trivial exercise, but a rough estimate is useful. 
We estimate the travel-time shift in the geometric optics limit as
the path integral of the fractional sound-speed perturbation 
(Equation 1 of \opencite{Hartlep2011})
weighted by the inverse of the background sound speed
in Model S.
For convenience, the path is chosen as purely horizontal through the center
of the perturbation. This procedure yields a travel-time increase of 23 seconds. 
 
\section{The Nominal Procedure and Results} \label{S-Nominal}      

Helioseismic holography (hereafter HH) is described extensively elsewhere 
(\opencite{Lindsey1997};\, \opencite{Chang1997};\, \opencite{Lindsey2000a},\citeyear{Lindsey2004b}).
For our purposes, it is useful to enumerate the data-analysis 
steps taken to define the ``nominal,'' or baseline, procedure. This provides
the context for investigating the sensitivity of the results to changes
in methodology discussed in Section \ref{S-Tests}.

\begin{figure}
\centerline{\includegraphics[width=1.0\textwidth]{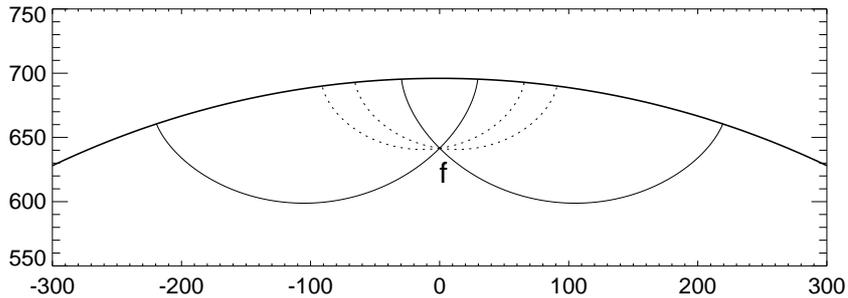}}
\caption{
Ray paths for $p$-modes converging on a focal point $f$ that is
54.4 Mm below the surface of the spherical domain of the
simulation (denoted by the thick
solid line). The thinner solid lines denote rays spanning $\pm 45^\circ$
from the horizontal direction. This is the nominal range of impact
angles for lateral-vantage helioseismic holography. In Section \ref{S-widths} we
perform HH with smaller ranges of impact angles. The dashed
lines indicate the ray paths for impact angles spanning $\pm 7.5^\circ$
from the horizontal direction. The scale is in Mm.
}
\label{fig.geom}
\end{figure}

The basic idea is to apply Green's functions to the solar oscillation
field at the surface of the Sun (or in this case, a simulation)
to estimate the amplitudes of incoming and outgoing waves at targets
(or ``focal points'') at or below the surface. 
In the ``lateral-vantage'' or ``deep-focus''
configuration of HH \cite{Lindsey2004b,Braun2004,Braun2008b},
travel-time perturbations are extracted
from the cross-covariances between these ``ingression'' and ``egression'' 
amplitudes with a focus below the surface (Figure~\ref{fig.geom}).

To establish some common notation, we define the
three-dimensional (3D) Fourier transform in time [$t$] and two spatial dimensions 
[$x, y$] of a function $A(x, y, t)$ as $\hat{A}(k_x,k_y,\omega)$ where
$k_x$ and $k_y$ are the horizontal wavenumber components and $\omega$ is the 
temporal frequency. We define
the Fourier transform in only the temporal dimension of $A$ as 
$\tilde{A}(x,y,\omega)$. 
With this, the steps involved in the data analysis
are listed below. 

\begin{enumerate}[i)]

\item The simulated surface-velocity data, 
provided in heliographic coordinates, are remapped 
onto a Postel projection $\Psi(x, y, t)$. The nominal
spacing of the Postel grid is $\delta x = \delta y = 8.54$ Mm 
(0.7$^\circ$), which is the original spacing of the velocity data in heliographic coordinates. The 
central tangent point $(x, y) = (0,0)$ is defined as 34.2 Mm (2.8$^\circ$) 
south of the location of the perturbation.

\item The 3D Fourier transform [$\hat{\Psi}(k_x,k_y,\omega)$] 
of the Postel-projected data is computed 
in both spatial
dimensions and in time. In the temporal-frequency domain, the data
within a bandpass of 2.5 and 5.5 mHz are extracted for further analysis.
We note that the simulation contains very little $p$-mode power
above 3.5 mHz.

\item\label{filter} A phase-speed filter is applied to $\hat{\Psi}$. 
The nominal method
employs filters that are Gaussian in the phase speed
[$w \equiv \omega/k$ (where $k^2 = {k_x}^2 + {k_y}^2$)] 
for each depth with peak phase speeds [$w_o$] and 
widths [$\delta w$]
specified in Table~\ref{T-modes}. 
In Section \ref{S-filters} we examine the sensitivity of the results to
variations in the form of the filter.

\item A set of depths is chosen (Table~\ref{T-modes}) and Green's functions
for both diverging [$G_+^P$] and converging [$G_-^P$] waves are
computed in the same Postel projected grid as the data \cite{Lindsey2000a}. 
The Green's functions are multiplied by spatial masks defining
a given pupil [$P$]. The nominal set of pupils represent 
quadrants (or ``arcs'') of annuli extending outward in
four directions and are denoted E, W, N, and S.
The annulus widths are determined by ray theory from the paths of acoustic 
modes diverging
from the subsurface focus point and spanning a range of ``impact angles'' 
$\pm 45^\circ$ from the
horizontal direction (see Figure~\ref{fig.geom}). 
In Section \ref{S-widths} we explore the sensitivity of the results to
narrower ranges of impact angles and 
in Section \ref{S-arcs} we employ different azimuthal extents of the pupil arcs.

\item\label{convolution} For each pupil quadrant [$P$], the ingression [$H_-^P$] 
and egression [$H_+^P$] amplitudes 
are estimated by convolutions of the data cube [$\Psi$] with $G_-^P$ and $G_+^P$, 
respectively, in both time and the two spatial coordinates.
This is performed using a plane-parallel approximation by 
the simple product of $\hat{G_{\pm}^P}$ and 
$\hat{\Psi}$ \cite{Lindsey2000a} in the three-dimensional Fourier domain. 
The validity and consequences of this approximation are explored in Section \ref{S-geom}. 

\item The cross-covariance functions between ingression and egression amplitudes
corresponding to opposite quadrants ({\it e.g.} E and W, N and S) are computed. 
The four resulting cross-covariance functions are summed.

\item\label{dt} Mean travel-time maps are determined from the sum of the four cross-covariance
functions. The nominal method uses the ``phase method'' \cite{Braun2000}. In
Section \ref{S-dt} we compare the phase method with results from fits of the
cross-covariances to Gabor wavelets. 

\item\label{flat} Maps of the mean travel-time shifts are determined from the residual
of the travel-time maps after subtracting a two-dimensional polynomial fit to a 
``quiet-Sun'' area excluding the perturbation.
As shown in Section \ref{S-geom} this procedure helps to remove the effects
of the plane-parallel approximation used in step~\ref{convolution}).

\end{enumerate}

Table~\ref{T-modes} shows the pupil ranges for each selected focus depth $z_f$, 
determined
from ray theory.
Also listed in the table are the range of mode degrees [$\ell$] at 3 mHz, 
sampled by the pupil, and the parameters for the 
Gaussian phase-speed filter (see Section \ref{S-filters}) at each depth. The highest value of $\ell$ at each depth 
represents waves propagating horizontally through the focal point 
while the lowest value indicates modes which propagate at impact
angles of $\pm 45^\circ$ from the horizontal direction (see Figure~\ref{fig.geom}).

\begin{table}
\caption{ Pupil sizes, modes, and filter parameters
}
\label{T-modes}
\begin{tabular}{ccccc}     % define the column alignment
                           % l: left, c: center, r: right
  \hline                   % horizontal line
$z_f$ & Pupil Radii & $\ell$ at 3 mHz & $w_0$ & $\delta w$ \\
  {[}Mm]    &    [Mm]         &      & [km s$^{-1}$] & [km s$^{-1}$]\\
  \hline
%1\tabnote{First table line.} & 02--Nov--97 & 16  & 24.1~ & -1.26 \\
 29.9 &  16.0\,--\,128 &  124\,--\,175 &   74 &   37 \\ 
 37.0 &  19.5\,--\,159 &  108\,--\,153 &   87 &   43 \\ 
 45.4 &  24.4\,--\,190 &   95\,--\,134 &   96 &   49 \\ 
 47.6 &  25.8\,--\,195 &   92\,--\,130 &  101 &   50 \\ 
 49.9 &  27.1\,--\,209 &   89\,--\,126 &  105 &   52 \\ 
 52.4 &  28.5\,--\,216 &   86\,--\,122 &  108 &   54 \\ 
 54.4 &  29.2\,--\,224 &   84\,--\,119 &  111 &   55 \\ 
 57.1 &  31.3\,--\,230 &   82\,--\,115 &  114 &   57 \\ 
 59.2 &  32.0\,--\,237 &   79\,--\,113 &  117 &   58 \\ 
 62.1 &  34.1\,--\,251 &   77\,--\,109 &  119 &   60 \\ 
 64.5 &  36.2\,--\,254 &   75\,--\,106 &  122 &   62 \\ 
 76.1 &  41.8\,--\,292 &   67\,--\, 95 &  140 &   69 \\ 
 87.9 &  48.0\,--\,327 &   60\,--\, 85 &  153 &   77 \\ 
  \hline
\end{tabular}
\end{table}

\begin{figure}
\centerline{\includegraphics[width=1.0\textwidth]{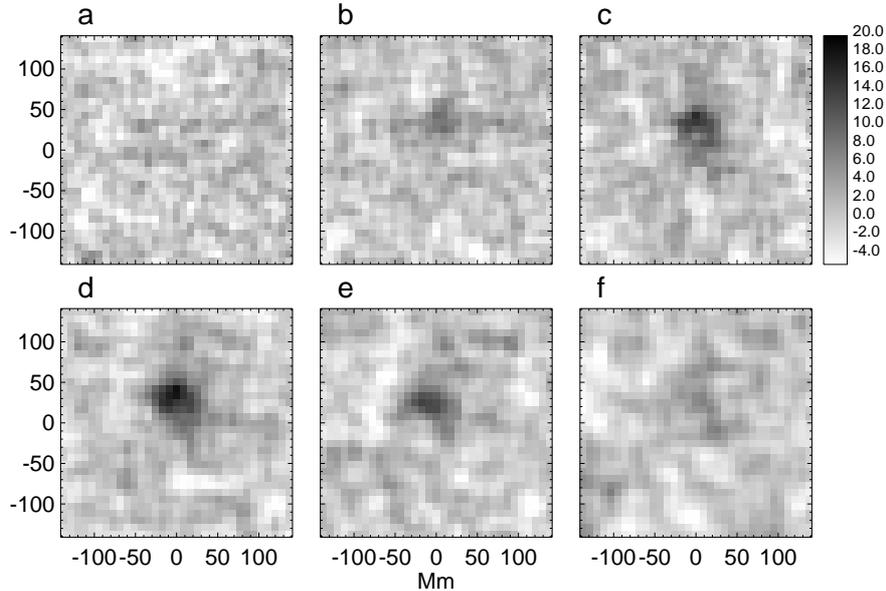}}
\caption{
Maps of the mean travel-time shift using the nominal methodology
of lateral-vantage HH as applied to the Hartlep {\it et al.} simulation and for
focus depths of: (a) 29.9 Mm, (b) 37.0 Mm, (c) 45.4 Mm,
(d) 54.4 Mm, and (e) 64.4 Mm. 
The greyscale indicates the travel-time shift in seconds.
}
\label{fig.nom2}
\end{figure}

Figure~\ref{fig.nom2}  shows maps of the travel-time shifts for a sample
of focus depths. The perturbation is clearly seen as an increase in travel-time shift
with a maximum of between 15 and 20 seconds at the expected horizontal position. 
Figure~\ref{fig.nom1}c shows the azimuthal averages of the travel-time shifts
for several focus depths while Figure~\ref{fig.nom1}d shows the variation of
the travel-time shift at the center of the perturbation (hereafter ``peak
travel-time shift'') with focus depth. It is clear that the horizontal
and vertical dependence of the travel-time shifts reasonably characterizes the shape
of the perturbation. 

We measure a background realization noise [$\sigma$] as
the standard deviation of the mean travel-time shifts within an
annulus spanning distances 111\,--\,195 Mm from the center of the Postel
projection. For the ``nominal'' maps
shown in Figure~\ref{fig.nom2}, $\sigma$ is about 2.1 seconds and does
not vary substantially with depth. 
We find that the background noise for maps made at different focus depths 
is correlated. For example, there is a correlation coefficient (measured
after excluding a region around the perturbation)
of 0.96 between maps made at 54.4 and 52.4 Mm, and a correlation of 0.56 
between maps at 54.4 and 47.6 Mm. 

\section{Tests of the Methodology} \label{S-Tests}      

\subsection{Tests of the Plane-parallel Approximation} \label{S-geom}      

In Section \ref{S-Nominal} step~\ref{convolution}), a convolution in time and
horizontal spatial coordinates between the Green's functions $G_{\pm}^P$
and the data $\Psi$ is computed in the Fourier domain under the 
assumption that the functions $G_{\pm}^P$ are invariant with respect to
translation in the Postel coordinate frame (this assumption has been termed
the ``plane-parallel'' approximation; \opencite{Lindsey2000a}; \opencite{Braun2008}).
The use of this approximation is highly desirable since it decreases 
computing time and resources by several orders of magnitude. Without its use,
for example, separate Green's functions for each target pixel 
would have to be computed, stored, accessed, and operated on with a 3D 
multiplication by the datacube in the computations.

\begin{figure}
\centerline{\includegraphics[width=1.0\textwidth]{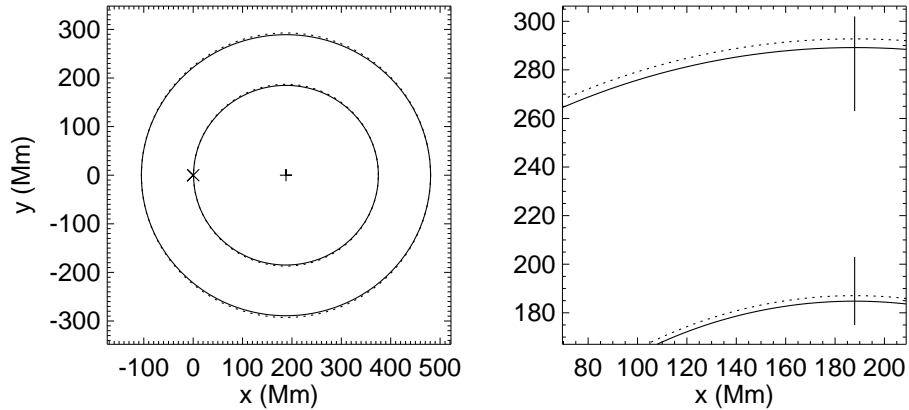}}
\caption{
Examples of
the distortion in wavefronts at the outer pupil boundaries centered on 
a target focus point (denoted by the plus sign) placed 188 Mm to the right of the center of the Postel
projection which is denoted by the $\times$ symbol. 
The larger and smaller dotted circles show wavefronts at the outer 
pupil boundaries, as projected onto the Postel frame,
for focus depths of 76.1  and 45.4 Mm respectively. The solid circles show
circular wavefronts as assumed in the plane-parallel approximation. 
The dotted and solid wavefronts coincide along the $x$-axis but deviate at
other places, with the maximum deviation occurring at the 
the top and bottom.  The deviations
are difficult to discern by eye in the left panel. 
The right panel shows a magnified
version of the upper part of the left panel.
The vertical line segments in the right panel indicate the length of the 
horizontal wavelengths of modes that propagate from the focus depth to the outer 
pupil boundaries.
}
\label{fig.pup2}
\end{figure}

\begin{figure}
\centerline{\includegraphics[width=1.0\textwidth]{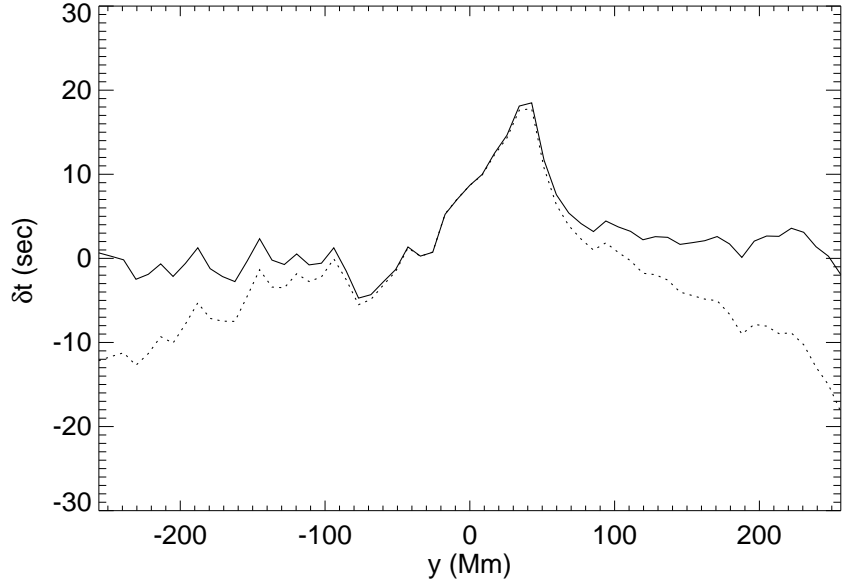}}
\caption{
Comparison of mean travel-time shifts with and without a polynomial subtraction
designed to remove the effects of the plane-parallel approximation. The dotted 
(solid) line is a vertical cut (at $x = 0$)  of the uncorrected (corrected) 
mean travel-time map at a focus depth of 54.4 Mm.
}
\label{fig.slice}
\end{figure}

A major result of this approximation is the introduction of a systematic 
bias in the mean travel-time shift, which is a function of the 
horizontal distance between 
the focus and the central tangent point of the Postel projection. The reason for
this bias is straightforward to understand: In the Postel (also known
as azimuthal equidistant)
projection, distances measured along any line intersecting the 
central tangent point (hereafter simply called the ``center'') are accurate, 
but distances between all other points 
differ from their true great-circle values. Thus, a locus of constant phase 
of waves propagating either away from or towards
the center is warped in the projected plane into an ellipse 
with the semi-minor axis aligned towards the center
(Figure~\ref{fig.pup2}). These wavefronts do not match 
the assumed circular wavefronts (and pupils) of the computed Green's functions. 
For the depths and pupil parameters 
listed in Table~\ref{T-modes} the distortion in distance is small
compared to the horizontal wavelengths of the modes. For example, in 
Figure~\ref{fig.pup2} are drawn wavefronts at the outer pupil boundaries 
(where the distortion is greatest) corresponding to focus positions
placed 188 Mm (15.5$^\circ$) to the right of the center and at 
depths of 45.4 and
76.1 Mm. The maximum distortion of the wavefronts for these depths
is 2.3 and 3.6 Mm respectively and
this is small compared to the horizontal wavelengths (30 and 40 Mm) of 
the modes considered.
However, the distortion in projected distances results in observable spurious
mean travel-time variations that vary with the azimuthal angle of propagation 
from the focus as well as the distance between the focus and 
the wavefront.  At the outer edge of
the pupils these spurious shifts can be as large as 20\,--\,30 seconds. 
However, the net travel-time shift as assessed over the
entire pupil is typically below ten seconds over tangent-point distances below 
200 Mm ({\it e.g.} Figure~\ref{fig.slice}).

To correct for this spatially varying bias, we fit and subtract a 
two-dimensional polynomial to the raw mean travel-time maps 
(Section \ref{S-Nominal} step~\ref{flat}). A circular mask excluding the
perturbation is applied before the polynomial fit. 
Figure~\ref{fig.slice} shows cuts
through a mean travel-time map with and without this correction.

\begin{figure}
\centerline{\includegraphics[width=1.0\textwidth]{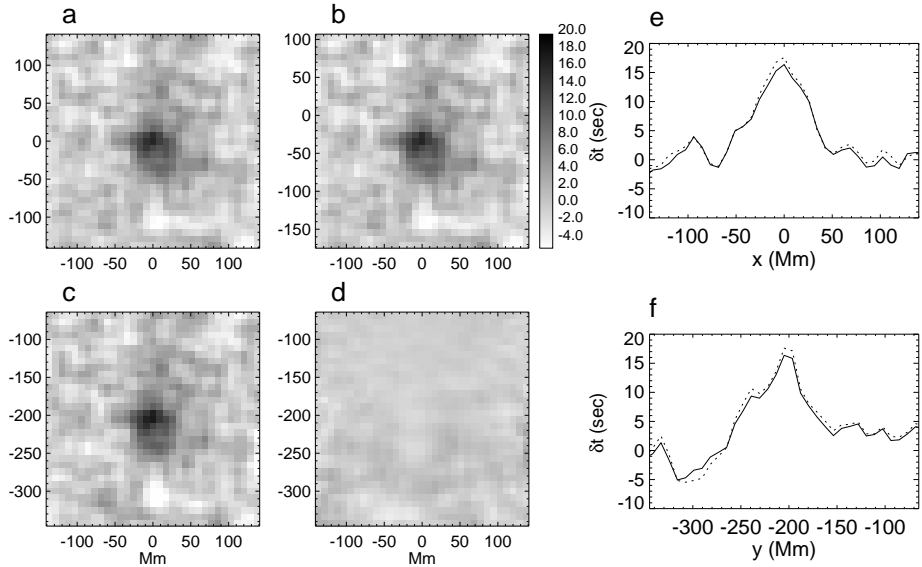}}
\caption{
Maps of the mean travel-time shifts at a focus depth of 54.4 Mm and with
the tangent point (center) of the Postel frame placed at the following locations:
(a) centered on the perturbation, (b) 34 Mm to the North of the perturbation, and
(c) 205 Mm to the North of the perturbation. 
(d) The difference between the maps shown in (c) and (a). 
The rightmost plots show horizontal (e) and vertical (f) cuts through
the center of the perturbation of map 
(a) shown as solid lines and map (c) shown as dotted lines.
}
\label{fig.planar}
\end{figure}

Since all of the distortions introduced by the plane-parallel approximation
increase with tangent-point distance, it is worthwhile to test the approximation
by computing travel-time shift maps with varying positions of the tangent point.
A similar test was performed by \inlinecite{Braun2012b} on solar observations, but
comparing only measurements of realization noise.
The present simulation provides a larger, isolated, signal that provides
a complementary target for this type of test. Figure~\ref{fig.planar} shows
that maps made using tangent points spaced 200 Mm apart show, after correction
for the bias discussed above, have residual differences on the order of a 
second. For smaller distances of approximately 20\,--\,30 Mm these residuals are well below
a tenth of a second.

\subsection{Comparisons of Travel-time Measurements} \label{S-dt}      

\begin{figure}
\centerline{\includegraphics[width=1.0\textwidth]{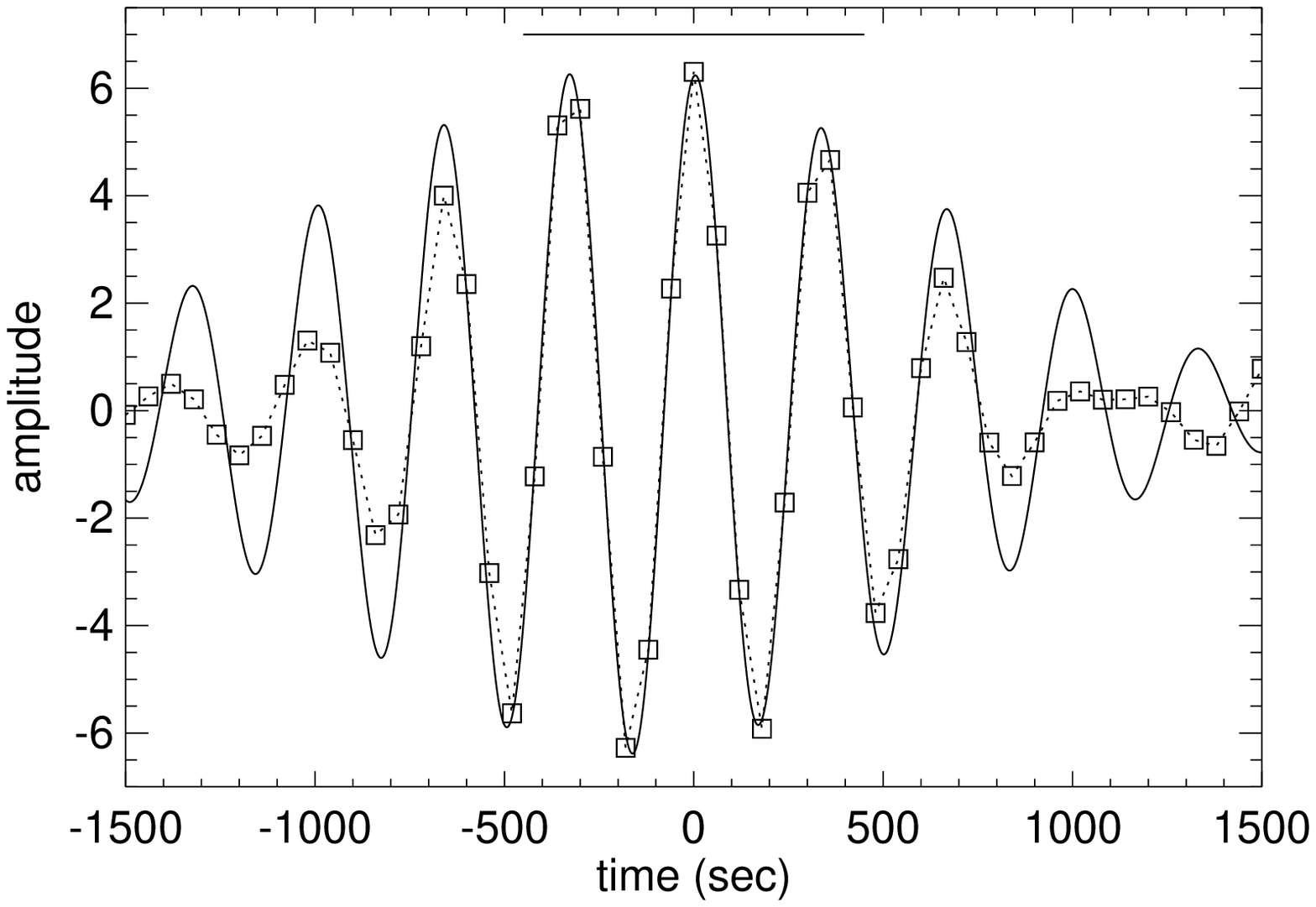}}
\caption{
The squares connected by dotted lines show a cross-covariance function between 
the ingression and egression amplitudes,
summed over the four opposite-quadrant pairs, for a single spatial location 
and a focus depth of 54.4 Mm. The solid curve represents a fit to the cross-covariance
function, sampled over a 14-minute window denoted by the horizontal line at the
top, to a Gabor wavelet (Equation~\ref{Eq-gabor}). The amplitude is in arbitrary units. 
}
\label{fig.corr}
\end{figure}

As we note in Section \ref{S-Nominal} step~\ref{dt}) the mean travel times are 
determined from the sum of the cross-covariance functions. 
In the nominal procedure, there are four cross-covariances of the form
\begin{equation}
\tilde{C}^{EW}(\mathbf r, z_f, \omega) = \tilde{H}_+^E({\mathbf r}, z_f, \omega) \tilde{H}_-^{W*}({\mathbf r}, z_f, \omega),
\label{Eq-EW_comp}
\end{equation}
where the asterisk denotes the complex conjugate, 
${\mathbf r} = (x, y)$, and we have also included the dependence on
focus depth $z_f$. The temporal Fourier transform of the sum,
\begin{equation}
\tilde{C} = \tilde{C}^{EW} + \tilde{C}^{NS} + \tilde{C}^{WE} + \tilde{C}^{SN},
\label{Eq-corr_sum}
\end{equation}
is used in the ``phase method'' \cite{Braun2000,Braun2008} to compute the mean travel time through 
\begin{equation}
\tau_{\mathrm {pm}} ({\bf r},z_f) = \arg\Big(\Big\langle \tilde{C}({\bf r}, z_f, \omega) \Big\rangle _{\Delta \omega}\Big) /  {\omega}_o,
\label{Eq-phase_meth}
\end{equation}
where the brackets indicate the average over the bandwidth $\Delta \omega$,
and ${\omega}_0$ is the mean frequency. The desired travel time shift [$\delta t$] is
obtained from $\tau_{\mathrm {pm}}$ by subtracting a 2D polynomial fit 
to a quiet-Sun region (step~\ref{flat}).
A typical summed cross-covariance function, transformed back to the temporal domain,
is shown in Figure~\ref{fig.corr}.

\begin{figure}
\centerline{\includegraphics[width=1.0\textwidth]{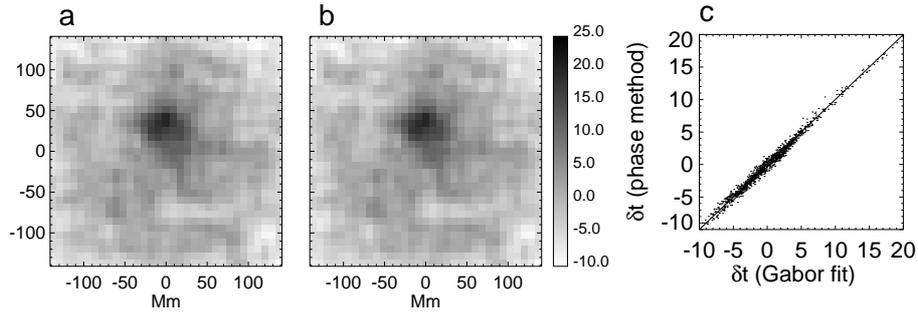}}
\caption{
Comparisons of maps of mean travel-time shifts
for a focus depth of 54.4 Mm obtained (a) using the nominal
method including the phase-method for extracting travel times from the cross
covariance functions and (b) using fits
to Gabor wavelets to the same cross-covariance functions in the temporal domain.
No corrections for the bias introduced by the plane-parallel approximation have
been performed here. Rather, a simple mean has been subtracted from each map.
(c) A scatter plot of the two maps, compared to a line with unit slope. 
}
\label{fig.fitcomp}
\end{figure}

An alternative method for extracting travel-time shifts is to fit the
cross-covariance function to a Gabor wavelet:
\begin{equation}
\label{Eq-gabor} g = A \cos(\omega_0[t-\tau_{\mathrm {gf}}])\exp\left( -\frac{1}{2}
  \left[{\frac{t-\tau_{\mathrm {en}}}{\sigma}}\right]^2 \right)
\end{equation}
where $A$, $\sigma$, and $\tau_{\mathrm {en}}$ are the amplitude, width,
and position of a Gaussian 
envelope,
$\omega_0$ is the mean frequency, and $\tau_{\mathrm {gf}}$ is the (phase) travel time
which is used instead of $\tau_{\mathrm {pm}}$ to determine the travel-time shift
$\delta t$.
We have applied MPFIT routines \cite{Markwardt2009}  to perform a 
non-linear least squares fitting 
of the summed cross-covariance functions to Gabor wavelets for a focus depth of 
54.4 Mm. The initial guesses of $\tau_{\mathrm {gf}}$ in the fits were 
based on the peak closest to $t=0$ 
of the cross-covariance function.
Figure~\ref{fig.corr} shows an example of the fit of a single cross-covariance.
Figure~\ref{fig.fitcomp} shows that 
there is remarkable agreement between the mean
travel time shifts as determined from the phase method and the Gabor fits.
We note that fine tuning the initial guesses based on the peaks of the
cross-covariance functions to the left (or right) of the central 
peak yields phase times $\tau_{\mathrm {gf}}$ which agree to within a 
fraction of a second of 
the times obtained using the
central peak minus (or plus) the period $2\pi/\omega_0$. 
Thus, due to statistical fluctuations in the mean frequency, maps made using
fits to these peaks are noisier than maps made using the central
peak.

\subsection{Sensitivity to Different Phase-speed Filters} \label{S-filters}      

\begin{figure}
\centerline{\includegraphics[width=1.0\textwidth]{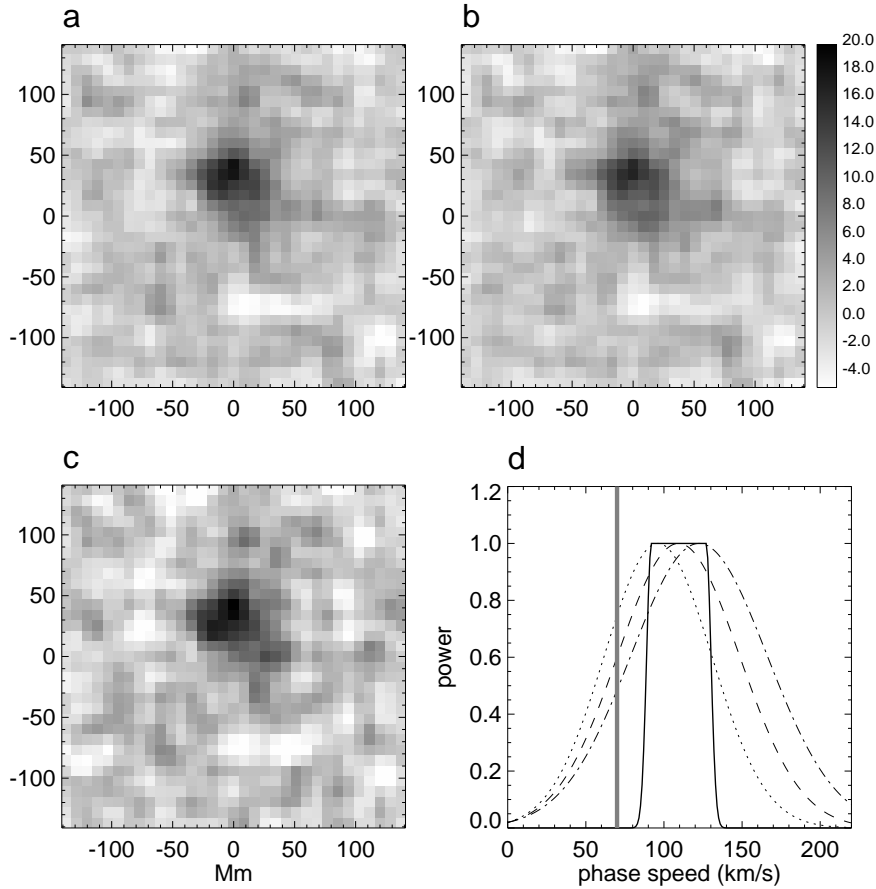}}
\caption{
Mean travel-time shift maps made for a focus
depth of 54.4 Mm and different phase-speed filtering: (a) a
Gaussian phase-speed filter, (b) a flat-top filter, and (c) no phase-speed
filtering.  (d) The square of the filter function 
for several filters used: dotted, dashed and dot--dash lines indicate
the nominal Gaussian filters corresponding to focus depths of 45., 54.4, and 64.5 Mm 
respectively. The solid line shows the flat-top filter used in this study. 
There is no $p$-mode power in the simulation to the left
of the vertical grey line. Thus, the tests here are not 
sensitive to differences between the filters
at these low phase speeds.
}
\label{fig.filter1}
\end{figure}

\begin{figure}
\centerline{\includegraphics[width=1.0\textwidth]{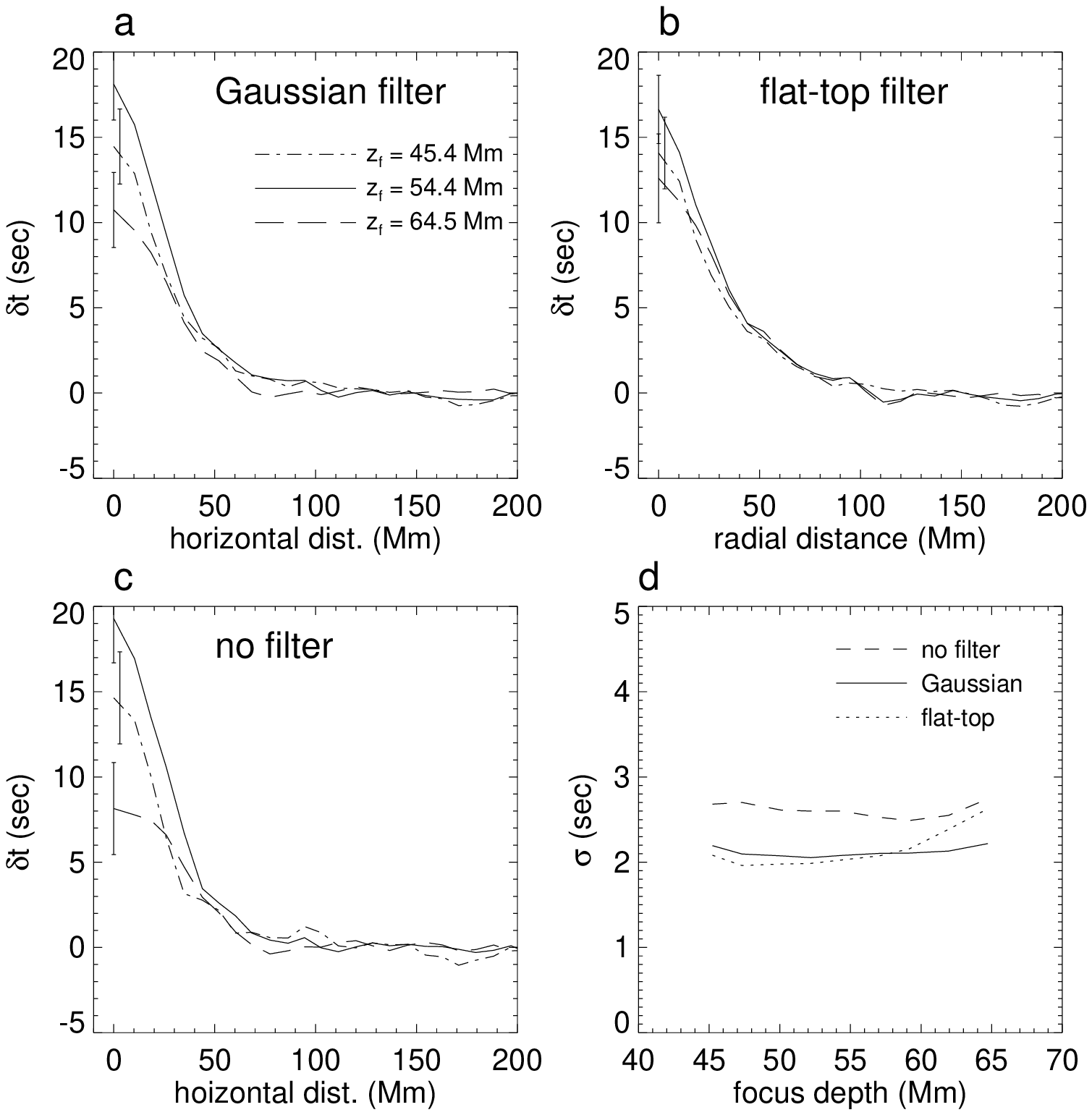}}
\caption{
Mean travel-time shifts for different focus depths, 
averaged in annuli centered on the perturbation (as in Figure~\ref{fig.nom1}c, 
using three different types of
phase-speed filtering: (a) the nominal Gaussian filters, (b) the flat-top
filter, and (c) no filtering.  The depths are
45.3 (dash--dot line), 54.4 (solid lines) and 64.5 Mm (long dashed line). 
Error bars represent the standard
deviation [$\sigma$] of the background realization noise in a region surrounding
the perturbation (see text). 
The variation of $\sigma$ with focus depth is shown in panel
(d) for the three cases: Gaussian (solid line), flat-top (dotted
line), and no filter (dashed line).
}
\label{fig.filter2}
\end{figure}

In Section \ref{S-Nominal} step~\ref{filter}), a phase-speed filter is applied to the Fourier transform
$\hat{\Psi}$ of the datacube. Phase-speed filters are widely used
in both time--distance helioseismology \cite{Duvall1997,Couvidat2006b}
and helioseismic holography \cite{Braun2006}.
The nominal procedure for lateral-vantage
HH uses Gaussian filters
\begin{equation}
\label{Eq-gauss} \phi = \exp\left( -\frac{1}{2}
  \left[{\frac{w-w_o}{\delta w}}\right]^2 \right)
\end{equation}
%\begin{equation}
%\label{Eq-gauss2} \phi = e^{{{ - \left( {w - w_o } \right)^2 } \mathord{\left/ {\vphantom {{ - \left( {w - w_o } \right)^2 } {2{\delta w}^2 }}} \right. \kern-\nulldelimiterspace} {2{\delta w}^2 }}}
%\end{equation}
with a peak phase speed [$w_o$] and width [$\delta w$]
such that square of the filter has values
of one and one-half at the highest and
lowest wavenumbers respectively at 3\,mHz as listed in Table~\ref{T-modes}.
It is recognized that the use of phase-speed filters reduces the noise contributed
by convective (non-wave) motions as well as from $p$-modes outside the 
range of desired phase speeds. 
Recently, \citeauthor{Ilonidis2012a} 
(\citeyear{Ilonidis2012b}, \citeyear{Ilonidis2012a}) 
claim that different types of filters
can affect the measured strength of subsurface signatures of emerging active
regions. We have compared results using the nominal Gaussian filter, results using
a ``flat-top'' filter similar to that employed by \inlinecite{Ilonidis2012b}, and
results using no phase-speed filter. Figures~\ref{fig.filter1} and \ref{fig.filter2}
show comparable peak travel-time shifts in the 
simulated perturbation between the three cases, although the flat-top filter
may be somewhat less sensitive to the variation with depth of the perturbation. 
This is also confirmed by computing the correlation coefficient between
maps for different depths. For example, travel-time shifts at focus
depths of 54.4 and 64.5 have a correlation of 0.45 using
the Gaussian filter, but 0.64 using the  
flat-top filter. These correlation coefficients were computed 
with the perturbation masked out,
so they measure correlations in the background realization noise.
Consistently higher correlation coefficients for
all of the flat-top filtered results over this depth range (45\,--\,65 Mm) are observed.
In general, the use of either filter produces somewhat less noise (as determined
from the standard deviation of the realization noise outside of the perturbation) 
than using no filter, as expected (Figure~\ref{fig.filter2}d). 

A restriction in the simulation to mode power below $\ell = 170$ means that
the tests performed here are not sensitive to variations in the filter properties
below $w = 70$ km s$^{-1}$. Nonetheless, our general findings are consistent with 
expectations based on experience analyzing solar data for lateral-vantage 
holography performed for similar focus depths.

\subsection{Sensitivity to Different Quadrant Widths} \label{S-widths}      

We explored the effect of changes to the range of impact angles of $p$ modes
interacting with the perturbation, by decreasing the pupil width from the
nominal values in Table~\ref{T-modes}. New pupil widths were computed
using ray theory for impact angle extrema of $\pm25^\circ$, $\pm15^\circ$, and
$\pm7.5^\circ$ at the focus depth of 54.4 Mm.  
Figure~\ref{fig.geom} shows rays corresponding to impact
angles of the nominal $\pm45^\circ$ and the smallest range, $\pm7.5^\circ$,
considered.

\begin{figure}
\centerline{\includegraphics[width=1.0\textwidth]{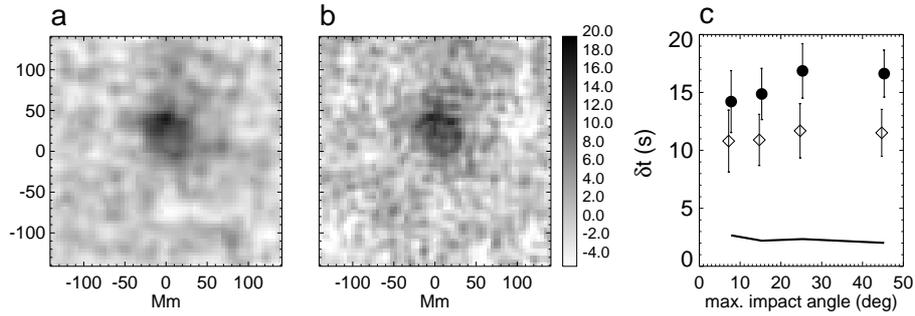}}
\caption{
Mean travel-time shift maps made for a focus
depth of 54.4 Mm and with different ranges of impact angle: (a)
$\pm45^\circ$ and  (b) $\pm7.5^\circ$. Both maps were made using the same
flat-top filter shown in Figure~\ref{fig.filter1}. (c) 
Measurements of the mean travel-time shift in the perturbation against
the maximum (absolute) impact angle. The filled circles show the peak
shift and the diamonds show the average shift within a 25 Mm radius.
Error bars denote the standard deviation of the realization noise, which is
also plotted as a solid line.
}
\label{fig.widths}
\end{figure}

Figure~\ref{fig.widths} shows that there is no substantial change in
the strength of the perturbation as the impact angle is changed,
within the uncertainty specified by the background realization noise.
This result is expected since the travel-time shifts due
to a simple sound-speed perturbation 
should not depend on impact angle. 
There is a slight increase of realization noise, which also appears to
take on a more fine-scale oscillatory  pattern, as the pupil quadrant
widths are decreased (Figure~\ref{fig.widths}b). This is likely
a diffraction (side-lobe) artifact due to the narrow pupil. 
The width of the pupils for these angles ($\pm7.5^\circ$) are smaller
than the horizontal wavelength of the modes.
To resolve this fine structure, the travel-time shift maps
shown in Figure~\ref{fig.widths} were made with a grid spacing of half of the
nominal value, resulting from an application of a Fourier interpolation 
of the original data.

\subsection{Sensitivity to Pupil Arc Size} \label{S-arcs}      

The advantage of using four quadrants to compute the 
ingression/egression cross-covariances derives primarily from 
the utility in making measurements sensitive to flows as well as perturbations producing mean (horizontal-direction-averaged)
travel-time shifts \cite{Gizon2005}. \citeauthor{Ilonidis2012a} (\citeyear{Ilonidis2012b}, \citeyear{Ilonidis2012a}) 
have proposed several refinements, for application to time--distance (hereafter TD) methods, for the detection of subsurface
signatures of emerging active regions. These include: i) dividing the annulus into a greater number of opposing arc pairs 
({\it i.e.} 6,\, 8,\, 10,\, 12, and 14 arcs), ii) making multiple measurements with different angular orientations of each set 
of arcs, and iii) combining all of the TD cross-covariances made with the different arcs and their orientations before
the determination of the travel times. There are four different orientations used for each arc configuration in
this scheme, as
each set of arcs is rotated one-quarter of the angular extent of an arc. 

\begin{figure}
\centerline{\includegraphics[width=1.0\textwidth]{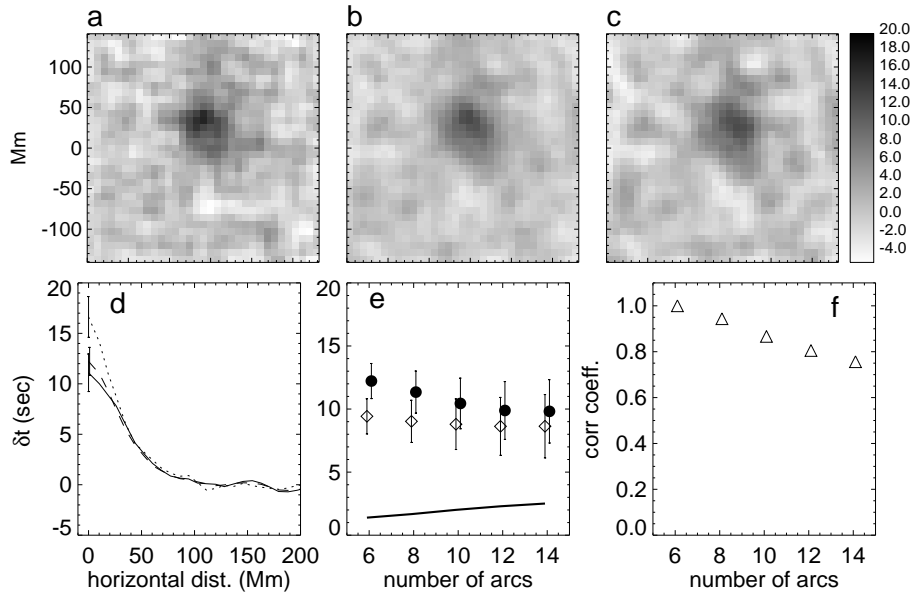}}
\caption{
Mean travel time shifts at a focus depth of 54.4 Mm from helioseismic 
holography using different pupil geometries: (a) 
the nominal method using a fixed set of four quadrant pupils, (b)  
using six arcs and four orientations, and (c) the combination of
6,\, 8,\, 10,\, 12, and 14 arc configurations with four orientations of each
configuration.
The flat-top filter shown in Figure~\ref{fig.filter1} is used.
(d) Azimuthal averages of the travel-time shift over annuli centered on
the perturbation for the three maps shown in the three top panels:
quadrants (dotted line), six arcs (dashed line), and combined 6\,--\,14 arcs
(solid line).
(e) Measurements of the mean travel-time shift 
in the perturbation and the background
realization noise $\sigma$ against
the number of pupil arcs used. The filled circles show the peak
shift and the diamonds show the average shift within a 25 Mm radius.
Error bars denote the standard deviation of the realization noise, which is
also plotted as a solid line.
(f) The correlation coefficient between travel-time shift maps made using six arcs and the other arc configurations.
}
\label{fig.arcs}
\end{figure}

We explore similar procedures for HH  using the simulation of
\inlinecite{Hartlep2011}. The results here complement the tests 
made for HH on Doppler observations obtained
with the MDI instrument by \inlinecite{Braun2012b}. 
Figure~\ref{fig.arcs} shows some of our
results of the measurements on the simulated perturbation. In general, the use of six arcs produces 
a weaker (by about 25\%) travel-time signature in the perturbation 
than using quadrants. 
A slight trend of a decreasing signal strength with the number 
of arcs from 6 to 14 is also observed (Figure~\ref{fig.arcs}e) although the 
net decrease is within the background noise.
The realization noise increases with the number of arcs used from about 1.4
seconds for the four orientations of the six-arc set to 2.5 seconds for the four orientations
of the 14-arc set. These can be compared with the 2 second noise measured using the nominal
quadrant method. These results are consistent with the increase in noise using smaller arcs
observed by \inlinecite{Braun2012b} using MDI data.
The map made from combining all cross-covariances from 
all pupil arc configurations and orientations has a realization noise of 1.9 seconds,
which is essentially identical to the nominal (quadrant) map.
It is significant that maps made with different 
arc lengths are highly correlated with each other (Figure~\ref{fig.arcs}f).

\section{Conclusions} \label{S-Conclusion} 
      
In summary, we find that helioseismic holography, as performed in the nominal
lateral-vantage configuration and using the plane-parallel 
approximation, is in conformance with expectations well suited for detecting and
characterizing subsurface sound-speed perturbations of the kind included
in the simulation of \inlinecite{Hartlep2011} at depths of at least 50 Mm. 
Suitable caution should be exercised since it is recognized that these
results follow from a single simulation, which may have different physics
from real solar perturbations.  Other limitations, such as the inclusion in
the simulation of only a subset (in both temporal frequencies and
wavenumbers) of known solar oscillations are noted. We believe that
in general, however, the viability of HH for the detection of subsurface
perturbations is substantially confirmed, particularly its ability to select
for analysis the relevant set of modes passing through a localized target 
below the solar surface. 

Furthermore, mindful of the caveats mentioned above, we find no 
evidence that the sensitivity of the procedure,
as assessed by the mean travel-time shift at the expected position of
the perturbation, is enhanced by the use of the flat-top filter or different
pupils as suggested in the critique by \inlinecite{Ilonidis2012b} 
of the results of \inlinecite{Braun2012b}. Specifically, the holographic
signatures are influenced little by the detailed profile of the phase-speed
filter, and, indeed, very little more by the lack of any such filter. 
In addition, we find that holography remains maximally sensitive when
applied with spatially extended pupils, as opposed to restricting or
partitioning it. The main effect of partitioning the pupil to smaller
arcs is, if anything, a reduction of the signature and the appearance
of diffraction effects.

Gabor-wavelet fitting can be applied to helioseismic holography as it is
with other time--distance techniques, and so this should not be regarded
as a discriminating qualification against it. In the case of the simulation,
the results are seen to be essentially identical to the phase-method and in
conformance with expected travel-time shifts given the size and amplitude
of the perturbation.

Our tests do not attempt to replicate
the time--distance procedures applied by \inlinecite{Ilonidis2011}. Thus we
draw no conclusion about the sensitivity of their own measurements to the 
changes to methodology that they advocate.  
In attempting to understand the discrepancies of the results between
\inlinecite{Ilonidis2011} and  \inlinecite{Braun2012b}, we can reasonably
infer that negative holography results suggest the suspected perturbation
is different than the simple sound-speed perturbation simulated by 
\inlinecite{Hartlep2011}. Furthermore, it would appear that the 
use of the plane-parallel approximation can be ruled out as a contributing
factor to the negative results of \inlinecite{Braun2012b}. 

It is possible that the physics of the suspected signatures may be such
that, unlike a simple sound-speed perturbation, the use of narrow pupils or
different filters may be critical.  
It is also suggested that the signatures may produce complicated changes to
the cross-covariance functions, perhaps due to unknown effects of magnetic
fields \cite{Ilonidis2012b}. 
Further tests need to be performed on the relevant data.
In our opinion, the possibility that the signatures of 
\inlinecite{Ilonidis2011} represent noise also needs to be considered.
We return to the point made by \inlinecite{Braun2012b} suggesting the
need for blind ``hare-and-hound'' tests as a minimal condition for the
signatures for the signals found by \inlinecite{Ilonidis2011} to be
established as pre-emergence signatures of deeply submerged magnetic fields. 
Tests with simulated data on artificial perturbations such as reported here
provide the critical context under which similar analysis of solar observations
may be understood. 
In general the results presented here provide confidence in helioseismic holography
as a useful method for probing submerged perturbations 
\cite{Leka2012,Birch2012}.

\begin{acks}
This work is supported by NASA Heliophysics program through contract NNH12CF23C, by the NASA SDO Science Center project
through contract NNH09CE41C, and by the Solar Terrestrial program of the National Science Foundation through
grant AGS-1127327. We thank Charlie Lindsey and an anonymous referee for useful comments. We are grateful to Thomas Hartlep for providing the 
data used in this study.
% The authors thank ... ({\it note the reduced point size})
\end{acks}

\bibliographystyle{spr-mp-sola}
\bibliography{db}

\end{article} 
\end{document}